\documentclass{emulateapj}
\usepackage{graphics}

\newcommand{\bi}{\begin{itemize}}
\newcommand{\ei}{\end{itemize}}
\newcommand{\be}{\begin{enumerate}}
\newcommand{\ee}{\end{enumerate}}
\newcommand{\nltt}{NLTT~11748}

\newcommand{\Mhe}{\ensuremath{M_{\rm He}}}
\newcommand{\Me}{\ensuremath{M_{\rm env}}}
\newcommand{\Rhe}{\ensuremath{R_{\rm He}}}
\newcommand{\zhe}{\ensuremath{\zeta_{\rm He}}}
\newcommand{\zrl}{\ensuremath{\zeta_{r_{\rm L}}}}
\newcommand{\Ma}{\ensuremath{M_{a}}}

\newcommand{\RL}{\ensuremath{R_{\rm L}}}
\newcommand{\Mdotyr}{\ensuremath{M_{\odot}\,{\rm yr}^{-1}}}
\newcommand{\lsim}{\lesssim}
\newcommand{\gsim}{\gtrsim}
\newcommand{\expnt}[2]{\ensuremath{#1 \times 10^{#2}}}   % scientific notation
\newcommand{\hmcnc}{HM~Cnc}
\newcommand{\rxj}{RXJ0806.3+1527}
\newcommand{\actaa}{Acta Astron.}
\begin{document}
\object[CSS 41177]
\object[SDSS J0106-1000]
\object[SDSS J0651+2844]
\object[KPD 1930+2752]
\title{Orbital Evolution of Compact White Dwarf Binaries}

\author{David L.~Kaplan\altaffilmark{1}, Lars Bildsten\altaffilmark{2,3},
  \&\ Justin D.~R.~Steinfadt\altaffilmark{3}} 
 
\altaffiltext{1}{Physics Department, University of Wisconsin -
  Milwaukee, Milwaukee WI 53211; kaplan@uwm.edu.}

\altaffiltext{2}{Kavli Institute for Theoretical Physics and
  Department of Physics, Kohn Hall, University of California, Santa
  Barbara, CA 93106; bildsten@kitp.ucsb.edu.}

\altaffiltext{3}{Department of Physics, Broida Hall, University of
  California, Santa Barbara, CA 93106; jdrsteinfadt@gmail.com.}

\slugcomment{ApJ, in press}

\begin{abstract} 

The new-found prevalence of extremely low mass (ELM, $\Mhe
<0.2\,M_\odot$) helium white dwarfs (WDs) in tight binaries with more
massive WDs has raised our interest in understanding the nature of
their mass transfer. Possessing small ($\Me \sim 10^{-3}\,M_\odot$)
but thick hydrogen envelopes, these objects have larger radii than
cold WDs and so initiate mass transfer of H-rich material at orbital
periods of 6--10 minutes. Building on the original work of D'Antona et
al., we confirm the $10^6\,$yr period of continued inspiral with mass
transfer of H-rich matter and highlight that the inspiraling
direct-impact double WD binary HM Cancri likely has an ELM WD donor.
The ELM WDs have less of a radius expansion under mass loss, thus
enabling a larger range of donor masses that can stably transfer
matter and become a He mass transferring AM CVn binary.  Even once in
the long-lived AM CVn mass transferring stage, these He WDs have
larger radii due to their higher entropy from the prolonged H burning
stage.

\end{abstract}


\section{Introduction} 

\begin{figure}
% plot_systems.m
\plotone{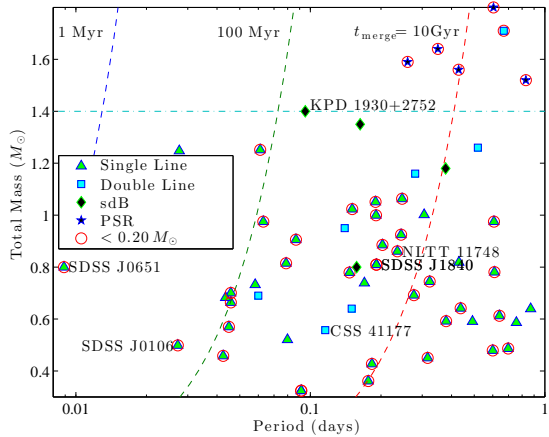}
\caption{\label{fig:wdmerg} The population of double WDs (triangles
  and squares, depending on whether one or two members of the binary
  have radial velocity measurements), sdB/WDs (filled diamonds), and
  pulsar/WDs (filled stars) with $P_{\rm orb}< $ day. Binaries to the
  left of the dashed lines will merge in less than 10 Gyr, 100 Myr, or
  1\,Myr due to gravitational wave losses. Binaries with a ELM He-core
  WD are circled.  Individual sources of interest are labeled:
  \nltt\ (the first eclipsing system), {CSS~41177}
  (the second eclipsing system; \citealt{pmg+11}), {SDSS~J0106$-$1000} and {SDSS~J0651+2844} (two recently discovered short period
  binaries; \citealt{kbk+11,bkh+11}), and {KPD~1930+2752} (a binary containing a massive WD and a
  sdB star with a total mass equal to the Chandrasekhar mass;
  \citealt*{mmn00}).  The line at the Chandrasekhar mass $M_{\rm
    Ch}=1.4\,M_{\odot}$ is where a merger could produce a type Ia
  supernova in the traditional scenario (but see \citealt*{vkcj10}).
  Also see \citet{kbap+12}.}
\end{figure}

Helium core white dwarfs (WDs) are made from $<2.0\,M_\odot$ stars
when stellar evolution is truncated before the He core reaches the
$\Mhe \approx 0.48\, M_\odot$ needed for the helium core flash. One
formation mechanism is significant mass loss due to stellar winds on
the red giant branch (RGB) that strips the H envelope \citep{ddro96},
typically leading to $\Mhe=0.4-0.48\, M_\odot$
(\citealt{hansen05,kbh+07}; \citealt*{ksp07}).  Another mechanism is a
common envelope induced by binary interactions (\citealt{il93};
\citealt*{mdd95}), making extremely low-mass (ELM) He WDs
($\Mhe<0.20\,M_{\odot}$ or so) when the interaction occurs at the base
of the RGB (see \citealt*{vkbk96}).  These ELM He WDs were first seen
as companions to millisecond pulsars \citep[e.g.,][]{bvkkv06} or in
high proper motion catalogs \citep{kvo+06,kv09}, but the advent of the
Sloan Digital Sky Survey \citep{elh+06} and other surveys revealed
many additional ELM WDs \citep[][and
  Figure~\ref{fig:wdmerg}]{ksp07,bmtl09,mbtl09,kbap+10,mgs+11,kvk10,sks+10,kbk+11,kbap+11,pmg+11,marsh11,bkh+11,kbh+11,vtk+11,kbap+12}.

ELM WDs were predicted to possess stably burning H envelopes ($\Me
\sim 10^{-3}-10^{-2}\, M_\odot$) that keep them bright for Gyrs
\citep{sarb02,pach07}, and this has certainly aided the recent
detections \citep{kbap+11,bkapk10}. Though identified by their
location in $\log g-T_{\rm eff}$ space, few systems have actually had
their radii measured with any precision.  Steinfadt et al.'s
(\citeyear{sks+10}) discovery of the eclipsing double WD binary system
\object[NLTT 11748]{\nltt}\ \citep{kv09} allowed for the first
geometric measurement of the radius of a ELM WD, finding $R\approx
0.04\,R_\odot$ for the $0.15\,M_{\odot}$ He WD, consistent with the
presence of a thick stably-burning hydrogen envelope (also see
\citealt*{kvv10}; \citealt{kapb+10b}).  Additional eclipsing systems
\citep{pmg+11,bkh+11} have led to even more constraints, although for
some of the more compact systems it is not clear if the radius is
truly the equilibrium radius of the WD or if it has been tidally
distorted.

Multiple common envelope phases are possible in the formation scenario
for ELM WDs, leading to ELM WDs in binaries with more massive
WDs. Many of these will come into contact within 10\,Gyr (Figure
\ref{fig:wdmerg}).  Indeed, a large number of the known double WD binaries
contain ELM WDs ($\Mhe <0.20 \, M_\odot$, circled points) with large
($>0.03 \, R_\odot$) radii indicative of a stable H burning shell
\citep{kvo+06,kbap+07,bmtl09,mbtl09,kbap+10,kvk10,mgs+11,bkh+11,kbk+11,vtk+11}.
As noted by \citet{davb+06}, since the time to burn the H envelope can
easily exceed the time to reach contact, many of these ELM WDs will
come into contact with the remaining H envelope. This raises the
possibility for many new phenomena that we begin to explore here.

The large radii of the ELM WDs means that Roche lobe overflow (RLO)
occurs at larger orbital periods than otherwise expected, so we start
in \S 2 by examining the behavior of the radius of the ELM WD as its H
envelope is transferred.  We follow in \S 3 by outlining the basics of
mass transfer and highlighting some of the new possibilities when ELM
WDs are donors. There is a more distinct possibility for thermally
stable burning of the accreted H and He on the accreting WD and the
initial contraction of the ELM WD to mass loss allows for more stable
mass transfer than that originally found for cold He WDs
\citep*{mns04}. We perform the full evolution calculations in \S 4,
highlighting the new phases of H mass transfer, the special behavior
near the period minimum, and the likelihood that the intruiging object
\object[HM Cnc]{HM Cancri} is one of
these systems.  We close in \S 5 by discussing the implications for AM
CVn evolution, and the remaining work needed to resolve the
thermonuclear outcomes for the accreted matter.

 \section{ELM Structure and Response to Mass Loss} 

The way in which the donor radius changes with mass loss
differentiates the evolution of these ELM donor binaries from the
earlier work of \citet{mns04} for cold He WDs, where the WD always
becomes larger as mass is removed. This is usually expressed as the
logarithmic derivative of radius with respect to mass $\zhe\equiv d\ln
\Rhe/d\ln \Mhe$ which is $\approx -1/3$  for cold,
degenerate matter.  However, the thick outer layer of non-degenerate
and stably-burning hydrogen in an ELM WD dramatically changes \zhe.

For this initial exploration, we use the models of \citet*{sba10} that
provide a range of hydrogen envelope masses \Me\ for each total
mass\footnote{Some calculations such as \citet{pach07} and
  \citet*{asb01} find that for masses greater than $\approx
  0.2\,M_\odot$, hydrogen will not burn stably but will instead
  undergo flashes. However, the precise boundary is not yet known and
  may be metallicity-dependent. For instance, the recently discovered
  binary \object[SDSS J065133.3+284423.3]{SDSS~J065133.33+284423.3}
  \citep{bkh+11} has an ELM WD with a high temperature and a radius
  indicative of burning even though it is somewhat more massive
  ($0.25M_\odot$) than the expected burning limit, although for this
  particular object the unknown contribution of tidal heating may
  contribute to its large size.  We limit our exploration to
  $\Mhe\leq0.20\,M_\odot$.}, \Mhe. These represent a large range of
ELM WD models ($0.125\, M_\odot \leq \Mhe \leq 0.20\, M_\odot$) that
adequately cover the possible envelope masses
($\Me=\expnt{(1-5)}{-3}\,M_\odot$) at the time of Roche lobe contact,
or equivalently have ages of 1--13\,Gyr; for an age of 1\,Gyr, the
expected envelope masses according to \citet{pach07} are
$\expnt{6}{-3}\,M_\odot$ ($\Mhe=0.16\,M_\odot$) to
$\expnt{7}{-4}\,M_\odot$ ($\Mhe=0.25\,M_\odot$), but the exact mapping
of \Me\ to age is not known in detail.  Instead, we verify the
starting models by noting that they span the observed range in surface
gravity and effective temperature seen in ELM WDs (e.g.,
\citealt{vtk+11}), with $T_{\rm eff}=8$,000--20,000\,K and
$\log(g)=5-7$.

\begin{figure}
% plot_radii.m
\plotone{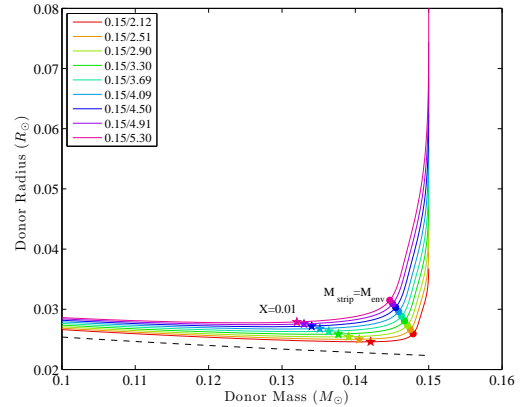}
\caption{Radii of $\Mhe=0.15 M_\odot$ He WDs as a function of
  remaining mass as the mass is stripped off.  The models are labeled
  by different starting envelope masses, $\Me$, in units of
  $10^{-3}\,M_\odot$.  We also plot the point where a mass equivalent
  to the envelope has been stripped (circles) and where the hydrogen
  mass fraction $X=0.01$ (stars).  Finally, the dashed line is
  Eggleton's zero-temperature mass radius relation
  \citep{vr88,mns04}.  Because of the diffusive tail, non-zero amounts
of hydrogen extend well inward of the point where the stripped mass is
equal to the envelope mass.}
\label{fig:radii}
\end{figure}

We need to find $\zhe$ as a function of the amount of mass that has
been lost, $\Delta\Mhe$, for each distinct initial model. These models
have the temperature, pressure and abundance profile of a
stably-burning pure H envelope in diffusive equilibrium with the He
core. The resulting transitional H/He layer has an abundance profile
set by chemical equilibrium in the changing electric field, as
described in \cite{sba10}.  As we show later, the thermal time at the
base of the H envelope is longer than the mass transfer timescale, so
that the material responds adiabatically to mass loss. To simulate
mass loss, we simply go to the mass coordinate $m=M_{\rm
  tot}-\Delta\Mhe$ in the initial model and force that mass element to
the surface by making its pressure artificially low (we went to
$P=10^{10}\,{\rm dyne\,cm}^{-2}$; in contrast, the pressure at the
H/He boundary is $10^{17}\,{\rm dyne\,cm}^{-2}$, while the pressure at
the core is $>10^{20}\,{\rm dyne\,cm}^{-2}$).

As the underlying fluid elements now have lower pressures, we conserve
their entropy by assuming that $TP^{-\gamma}$ is constant, where the
adiabatic index $\gamma$ was $\approx 2/5$ for the region in
question. The density is then computed from $T$ and $P$ following the
equation-of-state \citep[see][for details]{sba10}, and the composition
of each mass element is assumed to be the same as in the original
model.  Therefore, we need only integrate the pressure and radius with
respect to the mass coordinate.  This integration is done by shooting
from the core toward the surface and from the surface toward the core
and meeting at a mass coordinate in the middle (the $X=Y$ point, where
the hydrogen and helium mass fractions are equal).  Selected results for
0.15\,$M_\odot$ ELM WDs are shown in Figure~\ref{fig:radii}, where we
plot the radius as a function of the remaining mass along with the
zero-temperature (fully degenerate) model used by \citet{mns04}.  Two
things are obvious: (1) as the envelope is stripped, the slope $\zhe
\gg 1$ so the donor shrinks (as in \citealt{davb+06} for stable
burners), and the slope is still greater than the $-1/3$ value of the
zero-temperature model even once $X<0.01$; and (2) the radii are
always larger than the zero-temperature model, even when the envelope
has been stripped, because of the lower initial degeneracy.  In
our models the larger radii are consequences of both finite ages so
that even just passive cooling would result in warmer/larger WDs
(unlike in \citealt{mns04}) and stable H burning (unlike in
\citealt{dtwc07}).  The initial core temperatures at the onset of mass
transfer range from $\expnt{6.7}{6}\,$K for the
$\expnt{2.1}{-3}\,M_\odot$ envelope to $\expnt{1.4}{7}\,$K for the
$\expnt{5.3}{-3}\,M_\odot$, with higher temperatures resulting in higher
initial entropies \citep{db03}.

\begin{figure}
\plotone{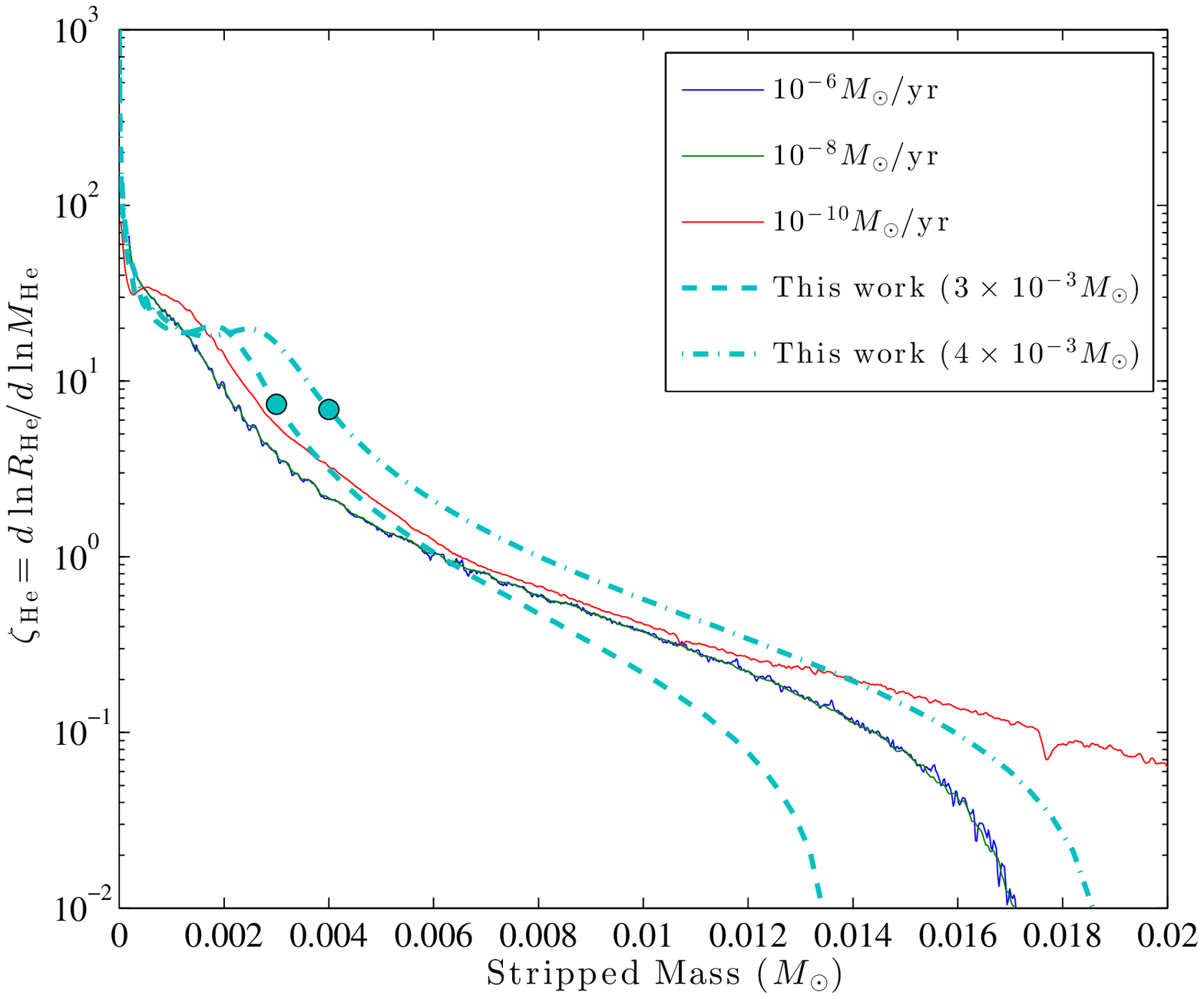}
\caption{Response of the initially stably burning donor to mass
  stripping $\zhe\equiv d \ln \Rhe / d \ln \Mhe$ as a function of
  stripped mass.  The initial models were for $\Mhe=0.15 M_\odot$ He WDs
  with either a $\Me=0.003M_\odot$  or $\Me=0.004M_\odot$ H envelope, undergoing
  stable hydrogen burning.  The dashed/dot-dashed lines show the
  results of this work using an adiabatic approximation to recalculate
  the stellar structure, with the filled circles indicating the
  envelope mass.  The solid curves show the dynamical results of
  \texttt{MESA} \citep{pbd+11} for a similar model (albeit with a
  lower envelope mass of $\Me=0.0021M_\odot$), using mass-stripping
  rates of $10^{-6}\,\Mdotyr$ (blue), $10^{-8}\,\Mdotyr$ (green), and
  $10^{-10}\,\Mdotyr$ (red).  Except for the transients at the
  beginning (which have to do with the outermost layers of the WD) all
  of the models generally agree for the first 10\% of the WD.  For the
  lowest accretion rate (not expected to be adiabatic) the model
  diverges after 0.015\,$M_\odot$.  In all cases, \zhe\ remains $>1$
  past twice the envelope mass.}
\label{fig:zeta}
\end{figure}

In Figure~\ref{fig:zeta} we compare our results for \zhe\ with those
computed using \texttt{Modules for Experiments in Stellar
  Astrophysics} (\texttt{MESA}; \citealt{pbd+11}).  For our starting
points, the models of \citet{sba10} only covered limited values of
\Mhe\ and \Me, as we used models that ``evolved'' from the starting
models.  Similarly, with \texttt{MESA} we had to use models that we
could construct out of suitable starting conditions, subject to mass
loss and diffusive equilibrium.  Therefore the comparison is not
exact: the \texttt{MESA} model has a total hydrogen mass of
$\expnt{2.1}{-3}\,M_\odot$, radius of 0.048$\,R_\odot$, and surface
temperature of 7150\,K, compared to ($\expnt{4.0}{-3}\,M_\odot$,
0.062$R_\odot$, 7160\,K) and ($\expnt{3.0}{-3}\,M_\odot$,
0.048$R_\odot$, 6500\,K), for our models.  We cannot simultaneously
match the envelope mass, radius, and surface temperature.  Even so,
the agreement for \zhe\ is good, and the model profiles (the pressure
and adiabatic index as functions of radius) agree well.  The main
apparent difference is the ``hump'' in \zhe\ around a stripped mass of
$\expnt{(2-3)}{-3}\,M_{\odot}$ visible in our models but not in the
\texttt{MESA} models.  This occurs slightly before the
hydrogen-to-helium transition and near the onset of degeneracy in the
original model, and given the different mass coordinates at which this
happens in our models versus the \texttt{MESA} models, the discrepancy
is not surprising.  We see that overall, the shape of $\zhe$ in this
adiabatic limit is constant in terms of $\Delta \Mhe/M_{\rm env}$.

Whether or not adiabatic evolution is a valid assumption depends on
the thermal timescale $\tau_{\rm th}$ at the base of the envelope,
which is where the luminosity is fixed and most of the excess radius
beyond the degenerate helium core is located. Mass-loss will be
adiabatic if $\Me/|\dot M|<\tau_{\rm th}$. For the models of
\citet{sba10}, $\tau_{\rm th}$ is typically a few Myr, so for
accretion that is faster than $\dot M_{\rm crit}\equiv \Me/\tau_{\rm
  th} \approx{\rm few}\times 10^{-9}\, M_\odot\,{\rm yr}^{-1}$, our
approximation is reasonable, and this regime encompasses most of the
evolution we discuss below. For lower mass-loss rates, we should
consider thermal evolution during the mass stripping, although at
these rates the amount of mass stripped is low enough that it does not
significantly affect our results.

\section{Mass Transfer Basics}

We start with a He WD donor with mass $\Mhe$ that is steadily losing
mass onto an accretor (typically a CO WD) with mass $\Ma$.  The total
mass is $M_t=\Ma+\Mhe$, while the binary orbital separation is $a$ and
the orbital angular momentum is $J=\Ma\Mhe(Ga/M_t)^{1/2}$. The mass
transfer rate is fixed by the rate of orbital angular momentum loss,
$\dot J$ via
\begin{equation}
{\dot J\over J}={\dot \Ma\over \Ma}+{\dot \Mhe\over \Mhe} + {\dot a\over
  2a}-{\dot M_t\over 2M_t}.
\end{equation} 
The accretor may have intermittent periods of unstable hydrogen
burning (classical novae, or CNe) that eject material from the binary;
therefore we cannot assume that mass transfer is conservative but must
instead track the mass lost by the binary, $\dot M_t$. We write this
as $\dot M_t=f\dot \Mhe$, so that $f=1$ means the accretor, on
average, keeps a constant mass (e.g., it ejects in each CN the amount
of matter that has accreted), and $f=0$ means that the accretor keeps
all the accreted mass.  We do not recalculate the accretor's radius as
it gains mass, as this is a small effect for the massive accreting WD.
If the CN were to excavate material from the accreting WD, then $f>1$,
although we do not explicitly consider this situation.  We write the
expression for $\dot J$ as
\begin{equation}
{\dot J\over J}={\dot \Mhe\over \Mhe}\left[1+(f-1){\Mhe\over
    \Ma}-f{\Mhe\over 2 M_t}\right]+{\dot a\over  2a},
\label{eqn:jdot}
\end{equation} 
allowing a connection between the mass transfer rate and the loss of orbital angular momentum. 

We always assume that the donor's radius \Rhe\ tracks that of the Roche lobe,
$\RL$, and use the simple \citet{pac67} formula for $\RL$, giving
\begin{eqnarray}
  {\dot J\over J}&=&{\dot \Mhe\over\Mhe}\left[1+{\zhe-\zrl\over
      2} +(f-1){\Mhe\over
      \Ma}-f{\Mhe\over 2 M_t}\right] \nonumber \\
 & \approx& {\dot \Mhe\over\Mhe}\left[{5\over 6}+{\zhe\over
      2}+(f-1){\Mhe\over
      \Ma}-f{\Mhe\over 3 M_t}\right].
  \label{eqn:Jdot}
\end{eqnarray}
Here, $\zrl\equiv d\ln r_{\rm L}/d\ln \Mhe$ is the derivative of the
Roche lobe radius in units of $a$, $r_{\rm L}=\RL/a$.  The Paczy\'nski
approximation is consistent to within about 5\% with the results using
the \citet{eggleton83} formula for $r_{\rm L}$ for the mass ratios
considered here.

At the onset of mass transfer, gravitational-wave losses set $\dot J$
to be \citep{ll75}:
\begin{equation}
\frac{\dot J_{\rm GR}}{J}=-\frac{32}{5}
\frac{G^3}{c^5}\frac{M_t\Mhe\Ma}{a^4}.
\label{eqn:gr}
\end{equation}
This remains true for wide systems, as the material transfers through
an accretion disk.  However, if there is not enough room for a disk
(based on Eqn.~6 of \citealt{npzvy01}) then material will impact
directly onto the accretor \citep{webbink84,mns04}.  In this case, the
angular momentum of the accreted material is lost to the accretor.
This angular momentum loss subtracts an additional
$\sqrt{r_h(1+\Mhe/\Ma)}$ from the quantity in the brackets in
Eqn.~(\ref{eqn:Jdot}), where $r_h$ is the equivalent radius of the
material orbiting the accretor (in units of $a$) as given by
\citet[][correcting for their inverted definition of the mass
  ratio]{vr88}.  In reality, all of the $J$ may not be lost but some
may be transferred back to the system via tidal coupling (as in
\citealt{mns04,fl12}), but for now we explore the limit where this
does not occur since the tidal timescales are largely unconstrained
for the accreting WD.

\subsection{Burning of the Accreted Material}
\label{sec:f}

The importance of knowing the value of $f$ is highlighted by Equation
(\ref{eqn:jdot}).  When $f=0$ (i.e., the transferred material stays on
the accretor), the mass ratio $\Mhe/\Ma$ decreases even more quickly,
reducing dynamical stability (\S~\ref{sec:stable}) and increasing
$|\dot \Mhe| $.  On the other hand, when material is lost from the
system (i.e., $f=1$), then $\Mhe/\Ma$ changes more slowly with a
corresponding reduction in $|\dot \Mhe| $.  It is the ability for the
accretor to burn the accreting material at the supplied rate that
determines $f$. Stable burning (i.e., burning the material at the
accreted rate in a thermally stable manner) implies $f=0$, whereas
unstable burning results in mass losing events (e.g., classical novae)
that drive $f\rightarrow 1$. Since no detailed calculations are
available that cover accreted material with our particular abundance
mixture (almost pure H going to nearly pure He); we base our current
calculations on \citet*{ppm82}; \citet{nskh07,sb07} for H burning and
\citet{it89} for pure He burning.

\begin{figure}
%\plotone{mdot_porb_0.8_fmassX_ages.eps}
\plotone{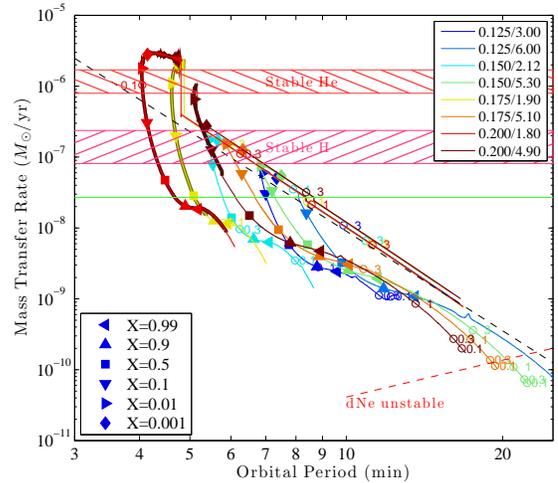}
\caption{Evolution of compact WD binaries involving an ELM He WD and a
  0.8\,$M_{\odot}$ CO WD.  The masses of the He WDs vary from
  0.125\,$M_{\odot}$ to 0.2\,$M_{\odot}$, and the envelope masses (in
  units of $10^{-3}M_\odot$) are also listed.  We plot the expected
  mass-transfer rate $\dot M$ from Roche-lobe overflow as a function
  of orbital period after the systems are brought into contact by GW
  emission.  The systems begin on the lower tracks, move toward
  shorter orbital periods and increasing $\dot M$, reach a period
  minimum, and then leave along the upper tracks.  The numbers show
  the ages of the systems (after contact) in Myr.  The thick regions
  are where the accretion stream hits the CO WD directly, instead of
  forming an accretion disk.  The tracks are additionally labeled with
  symbols that show the hydrogen fraction $X$ of the accreted
  material, changing from 99\% down to 0.1\%.  The black dashed line
  is Eqn.~\ref{eqn:mdotporb}.  The regions where stable H
  \citep{nskh07} or He \citep{it89} burning are possible are shown by
  the hatched regions, although we use the green horizontal line for
  the lower H-burning limit to account for reduced metallicity.
  Finally, the red dashed line shows where accretion is unstable to
  dwarf novae (dNe) for pure hydrogen, based on \citet{shafter92}.}
\label{fig:mdot}
\end{figure}

The transferred material at the onset of mass transfer is certainly
pure hydrogen. If that material does not gain any carbon or oxygen
contamination from the accreting WD, then it can only burn via the
pp-cycle. Though briefly considered by \citet{ppm82}, no recent
calculations have been done to explore the thermal stability of such a
pure H layer. \citet{sb07} performed stability calculations for very
low metallicity, going down to $10^{-4}$ Solar metallicity. They found
that the burning material was more stable, reducing the minimum
accretion rate where stable burning can occur by over a factor of
three from the limits derived by \citet[][Eq.~5]{nskh07} for Solar
metallicity (shown by the the red banded region in
Fig.~\ref{fig:mdot}). We therefore assume stable hydrogen burning
when $\dot M>10^{-7}(M_a/M_\odot-0.5357)\,M_{\odot}\,{\rm yr}^{-1}$
(the green horizontal line in Fig.~\ref{fig:mdot}), and assume that
all the accreted mass remains on the accretor ($f=0$). For rates lower
than this, we assume that the H burning is unstable, leading to Roche
lobe overflowing nuclear flashes and $f=1$.  However, very little mass
is transferred at these low rates, leading to a small impact on the
outcome.

The accretion rates needed for stable pure He burning are much higher
than those for hydrogen.  Just as in H burning, the stability arises
due to the increasing importance of radiation pressure and shell
thickening, both occurring as the burning luminosity approaches the
Eddington limit \citep{ty96,sb07}.  Most thoroughly discussed in
\citet{it89}, this rate depends on the WD core mass, and, for the
evolution of these binaries, allows for stable He burning only near
the period minimum \citep{ty96}.  This results in a narrow range of
rates (typically only a factor of $\sim 2$, see the hatched region in
Fig.~\ref{fig:mdot}) for which a WD can stably burn helium. This
demands two comparisons for our calculations. First, a tracking of
when the accretion rate is in the stable regime and for that duration,
a confirmation that enough mass is transferred to build the steady
state burning model. Second, when $\dot M$ reaches the upper limit of
the stable burning regime (meaning that the radius of the accretor
exceeds its Roche radius), we must account for the excess transferred
matter that cannot be stably burned by the accretor. For this
calculation, we simply assume (as is commonly done; \citealt{nskh07})
that that matter is lost by the binary.  Hence, even in the stable
regime we will sometimes set $f$ to a value that is different than
zero.\footnote{In contrast to the majority of our numerical
  integrations where we use an adaptive step size $4^{\rm th}/5^{\rm
    th}$-order Runge-Kutta integrator, updating the value of $f$ is
  done so as to keep the rate of accreted material for the last
  iteration in the stable regime --- no higher order derivative is
  used --- which leads to the apparent numerical noise in
  Figure~\ref{fig:mdot}.  This does not affect the outcome.}  We do
this so that $|\dot \Mhe|$ does not exceed the limit for stable He
burning (essentially the Eddington limit).  When these conditions
occur in our evolutionary scenarios, its almost always the case that
$X\approx 0$. Eventually, as the orbit widens, $|\dot \Mhe|$ decreases
below the lower limit for stable He burning, and we set $f=1$ again.

\subsection{Enhanced Stability of Mass Transfer from an ELM}
\label{sec:stable}
Mass transfer is dynamically stable when the term in the brackets in
Eqn.~(\ref{eqn:Jdot}) is positive for $f=0$.  If it is negative, the
stellar radius will be larger than the Roche radius after mass
transfer, leading to a dynamical instability and  merger.
The form of this condition depends on whether the angular momentum of the
material which leaves the donor can get back into the orbital angular
momentum (as assumed in disk accretion) or becomes ``lost" in the
accretor during direct impact. Accretion via a disk is stable when
\begin{equation}
\frac{\Mhe}{\Ma} < \frac{5}{6}+\frac{\zhe}{2}.
\label{eqn:stable}
\end{equation}
For a cold WD, $\Rhe\propto \Mhe^{-1/3}$, leading to
$\zhe=-1/3$. This would imply stability if ${\Mhe}/{\Ma}<2/3$, the
traditional criterion for stability. The
additional angular momentum loss allowed by direct-impact
changes the criterion to:
\begin{equation}
\frac{\Mhe}{\Ma} < \frac{5}{6} + \frac{\zhe}{2} -
\sqrt{\left(1+\frac{\Mhe}{\Ma}\right)r_h}.
\label{eqn:stabledirect}
\end{equation}
This substantially reduces the allowed mass ratios for initially stable mass
transfer \citep{mns04} especially for cold WDs, or 
those with a finite entropy \citep{dtwc07} but no burning.

However, as we show in Figure~\ref{fig:zeta}, He WDs with thick,
massive hydrogen layers have $\zhe \gg 1$ for stripped masses
$\Delta\Mhe/\Mhe\lsim 10$\%. \citet{davb+06} were the first to explore
the possibly different evolutions this allows for.  Just from
Eqn.~(\ref{eqn:stable}), $\Mhe<\Ma$ mass transfer will be dynamically
stable as long as there is an accretion disk.  Even without a disk
(using Eqn.~\ref{eqn:stabledirect}), the additional term
$\sqrt{(1+\Mhe/\Ma)r_h}$ is roughly $\approx 0.5$, so as long as
$\zhe$ is positive we still have dynamical stability. Hence, for
almost all the mass ratios considered here, dynamical instability is
not possible for disk-fed accretion and could only happen for
direct-impact accretion when $\Mhe/\Ma \gsim 1/4$ (assuming a typical
$\zhe=-1/3$ once the hydrogen envelope has been fully stripped). Tidal
interactions could work to stabilize these few unstable mass transfer
cases \citep[as in][]{mns04}, but that is beyond the scope of our
work.

\section{Evolution Including Mass Transfer from an ELM WD}

Just as in the conventional double WD binaries, the lower mass ELM WD
will be driven into contact by the loss of angular momentum from
gravitational waves. For the more massive companion, we presume cold
CO WDs with masses 0.7--1.0\,$M_\odot$ (we do not discuss masses
$<0.7\,M_\odot$, as those are not covered by the calculations of
\citealt{nskh07}).  We neglect any tidal heating of the ELM WD prior
to contact, and simply assume, as in the previous section, that it has
a thick H envelope that is stably burning. We start by discussing the
long period of stable mass transfer of the overlying H shell, then
discuss the behavior near the period minimum and close with a brief
discussion of the outgoing He mass transfer phase; which is similar to
that previously discussed in the literature.

\subsection{Mass Transfer Phase up to the Period Minimum} 

The larger stellar radius leads to the onset of mass transfer at
orbital periods of $\sim 10$ minutes, much longer than expected for a
thin-H-shell cooling WD. Since $\zhe\gg1$, the donor shrinks under
mass loss, so the mass transfer is stable and secularly driven at the
rate implied by equation (\ref{eqn:Jdot}) when angular momentum loss
is fixed by gravitational radiation (see Eqn.~\ref{eqn:gr}).  We
explore a range of ELM WDs with masses $\Mhe/M_\odot=$0.125--0.20 and
envelopes of $\Me=\expnt{1.5}{-3}$ to $\expnt{5.1}{-3}\,M_{\odot}$,
although a single donor mass only has part of the range.  Not all of
these are physically realistic: some higher-mass WDs may not have
stable burning, or some combinations of envelope and total mass may
not be reachable in a Hubble time (see discussions in \citealt{pach07}
and \citealt{sba10}).

\begin{figure}
%\plotone{early_evolution.eps}
\plotone{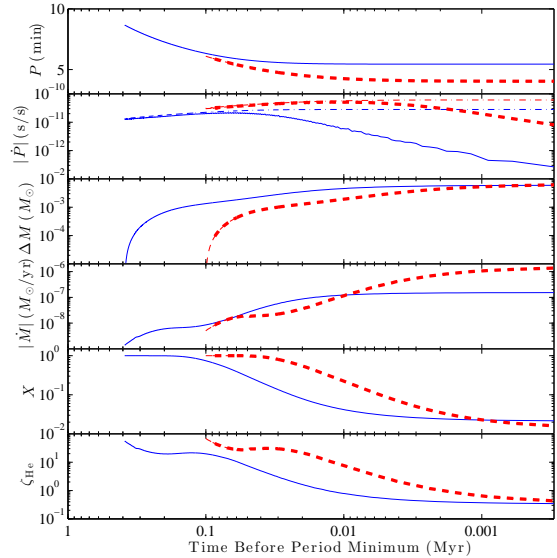}
\caption{Evolution of compact WD binaries involving an ELM He WD and a
  $\Ma=0.8M_{\odot}$ CO WD.  The masses of the He WDs are
  $\Mhe=0.15M_{\odot}$ (blue solid line) and $\Mhe=0.2M_{\odot}$ (red dashed
  line), with envelope masses of $\Me=\expnt{2.1}{-3}M_\odot$ and
  $\Me=\expnt{1.8}{-3}\,M_\odot$, respectively.  We plot (top to bottom)
  the orbital period, the orbital period derivative, the transferred
  mass, the accretion rate, the hydrogen mass fraction, and \zhe.  The
  period of direct-impact accretion for the $\Mhe=0.2M_{\odot}$ WD is
  indicated by the thick dashed line. In the $\dot P$ panel, the red and blue
  dot-dashed lines are where the loss of angular momentum is entirely
  given by Eqn.~(\ref{eqn:gr}).}
\label{fig:early}
\end{figure}

We start our calculation when mass transfer begins (also see
\citealt{davb+06}), typically at orbital periods of 6--10\,minutes.
Figure~\ref{fig:early} shows the resulting evolution up to the period
minimum for two cases with $\Mhe=0.15\,M_{\odot}$ and
0.2\,$M_{\odot}$.  For these cases, the accretion rate, $\dot M$, is
in the range where our adiabatic approximation is
valid.\footnote{Other cases we show in a Figure~\ref{fig:mdot} that
  come into contact at much wider orbital periods violate this
  inequality, but in those cases the amount of mass transferred at
  this low rate is small enough as to not qualitatively change the
  results.}  The increase in $|\dot M|$ as the orbital period shrinks
is evident. At the earliest part of the onset of accretion, the
orbital $\dot P$ (second panel) is nearly identical to that calculated
(dot-dashed lines in the second panel) from the loss of angular
momentum from two orbiting point masses. Namely, if measured at this
stage, the system would appear to have an orbital period change
consistent with gravitational inspiral \citep[][also see
  Fig.~\ref{fig:pdot}]{davb+06}.  During this initial evolution we
have $\zhe \gg \zrl$, and we find from Eqns.~(\ref{eqn:Jdot}) and
(\ref{eqn:gr}) that
\begin{equation}
|\dot \Mhe|_{\rm in}\approx \frac{64\Mhe^2\Ma}{5c^5\zhe}
\left(\frac{256\pi^8 G^5}{M_t P^8}\right)^{1/3}, 
\end{equation}
highlighting the need to calculate the considerable evolution in \zhe\ during this period
(Fig.~\ref{fig:early}) to find $\dot \Mhe$. 

This evolution is a new mode of H mass transfer onto a WD that has not
been previously explored over a wide range of initial masses with
semi-analytic models. The long ($\sim\,$Gyr) life of the ELM WD prior
to contact has allowed for complete diffusive equilibrium so that the
transferred material is pure hydrogen at the start, with an increasing
amount of helium at later times as the stripping reveals the
underlying He WD core (5th panel in Figure~\ref{fig:early}).  The
amount of mass that has been stripped (third panel down) is simply
what's needed to keep the stellar radius equal to the Roche radius as
the donor moves inward.  These are often but not always disk accretors
but compact enough that the thermal disk instability that gives rise
to dwarf novae is suppressed (systems below the dashed line on
Figures~\ref{fig:mdot} and \ref{fig:mdot2} would undergo dwarf novae
instabilities).

The thermonuclear outcomes during this phase remain uncertain
until a time dependent accretion calculation has been performed with
the changing $\dot M$ and $X$ of Figure~\ref{fig:early}.
 However, as we discussed in \S~\ref{sec:f}, if H mass transfer leads to 
stable burning, the system's luminosity would increase to  $L=Q_{\rm
  nuc}\dot M$.

\subsection{Near the Period Minimum and The Ultra-Compact Binary \hmcnc} 

\begin{figure}
%\plotone{pdot_porb_0.8_fmassX.eps}
\plotone{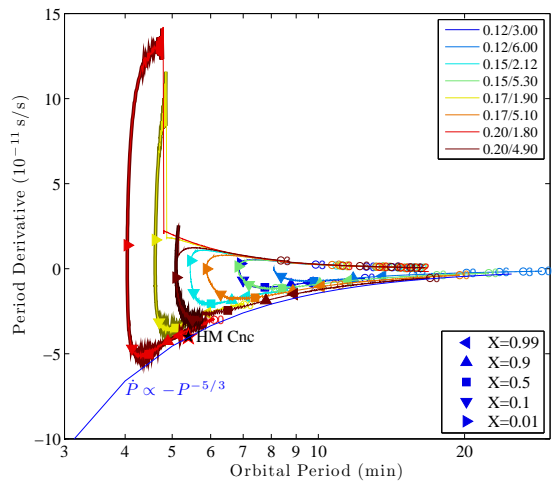}
\caption{Period derivative vs.\ orbital period for a compact WD binaries involving an ELM He WD
 and a 0.8\,$M_{\odot}$ CO WD.  The different donor models are as in
  Figure~\ref{fig:mdot}.  We also the expected $\dot P\propto
  -P^{-5/3}$ behavior along the incoming (lower) track
  (Eqn.~\ref{eqn:pdot}) as expected from  gravitational inspiral.  We
  also show the observed properties of \hmcnc.}
\label{fig:pdot}
\end{figure}

\begin{figure*}
%\centerline{\includegraphics[width=0.5\textwidth]{mdot_porb_0.7_fmassX.eps}\includegraphics[width=0.5\textwidth]{mdot_porb_0.8_fmassX.eps}}
%\centerline{\includegraphics[width=0.5\textwidth]{mdot_porb_0.9_fmassX.eps}\includegraphics[width=0.5\textwidth]{mdot_porb_1.0_fmassX.eps}}
\centerline{\includegraphics[width=0.5\textwidth]{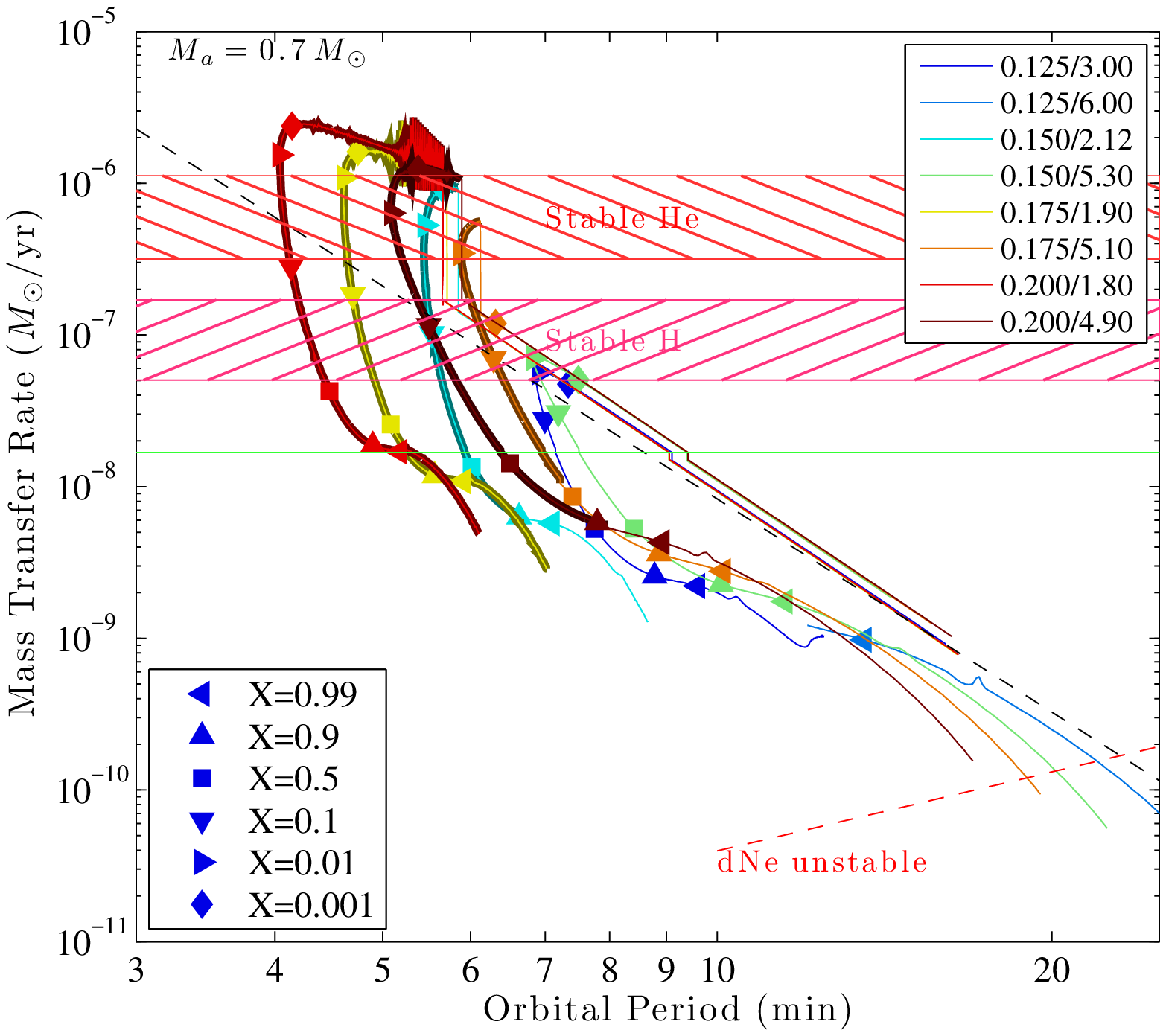}\includegraphics[width=0.5\textwidth]{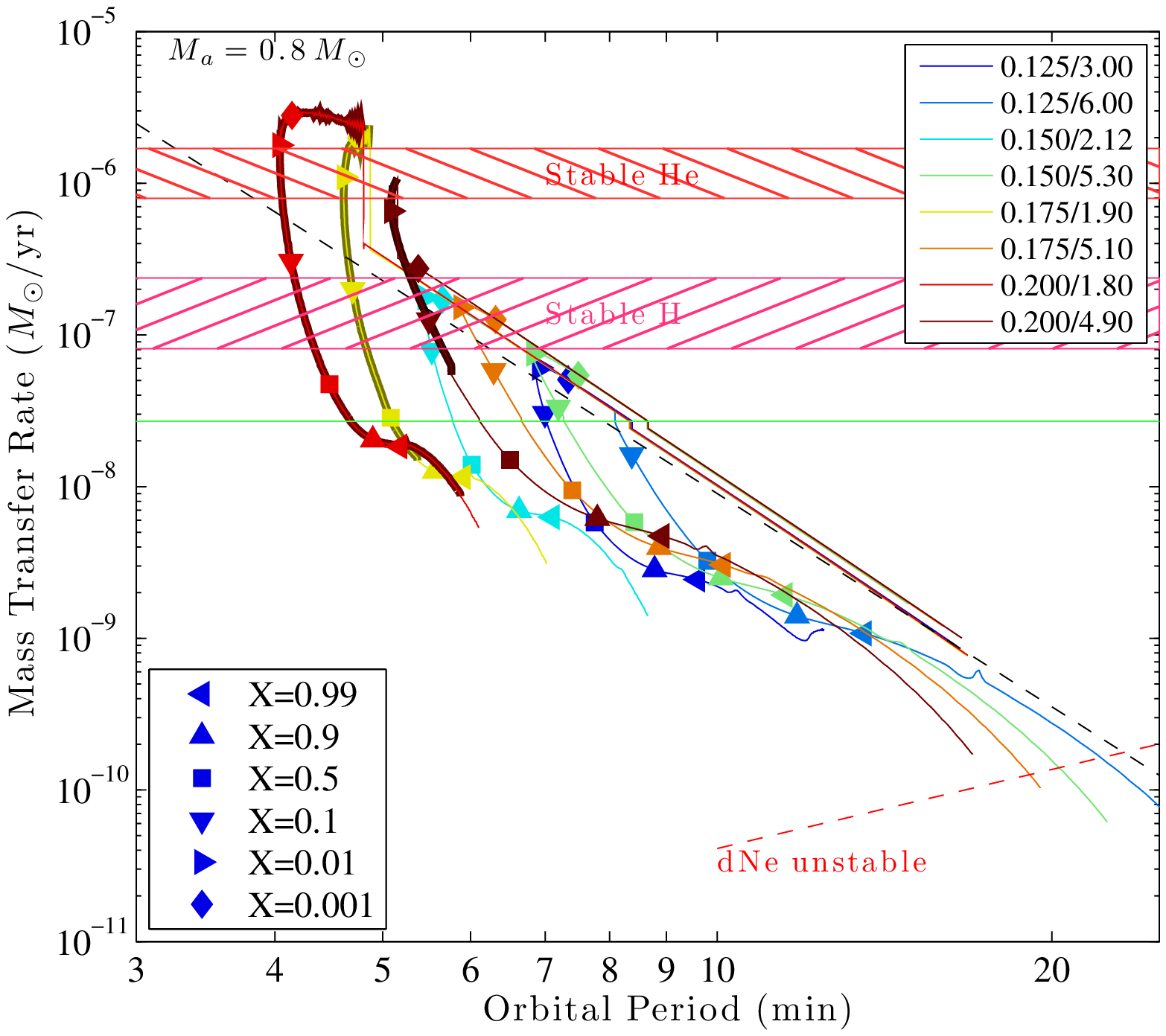}}
\centerline{\includegraphics[width=0.5\textwidth]{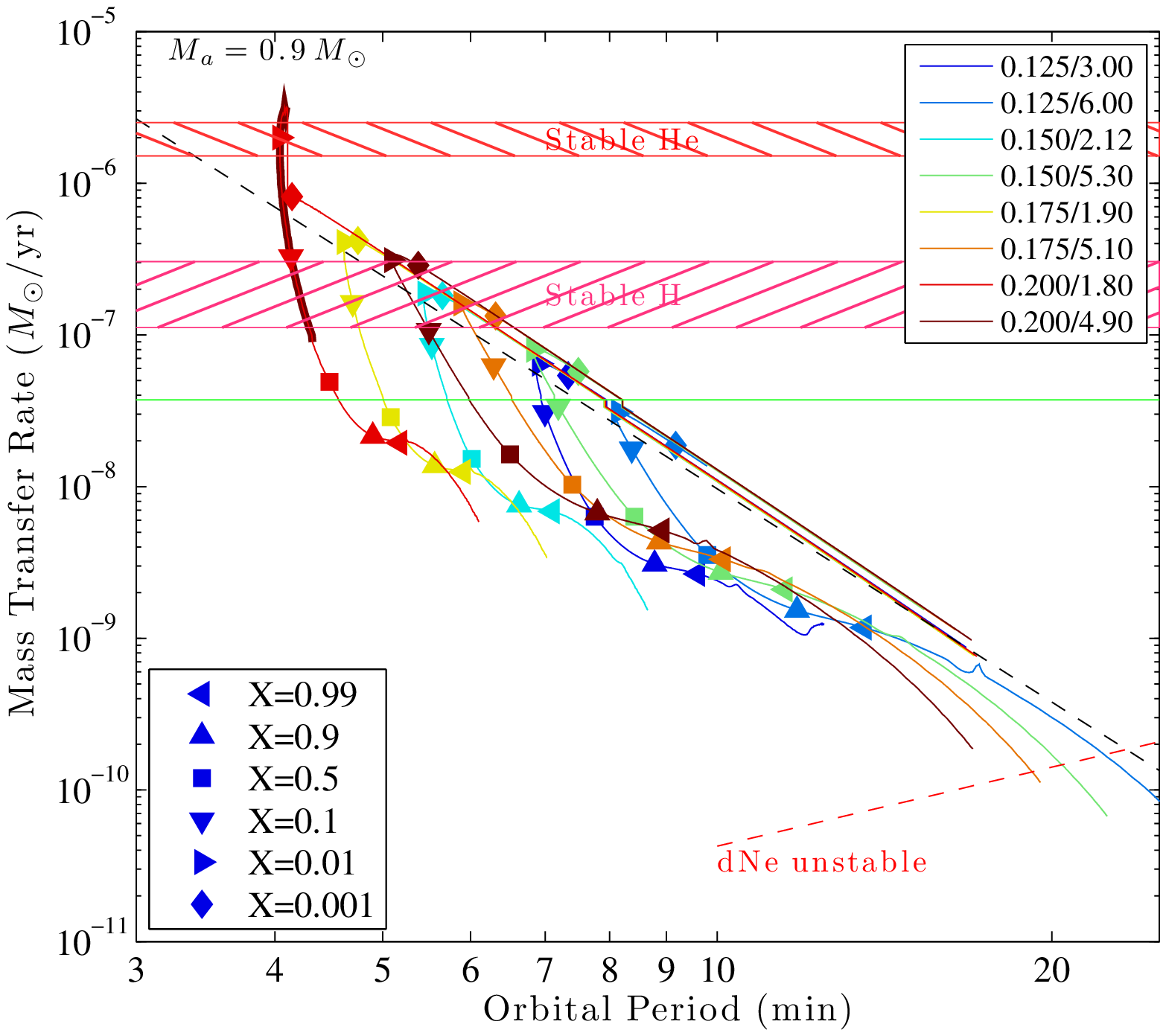}\includegraphics[width=0.5\textwidth]{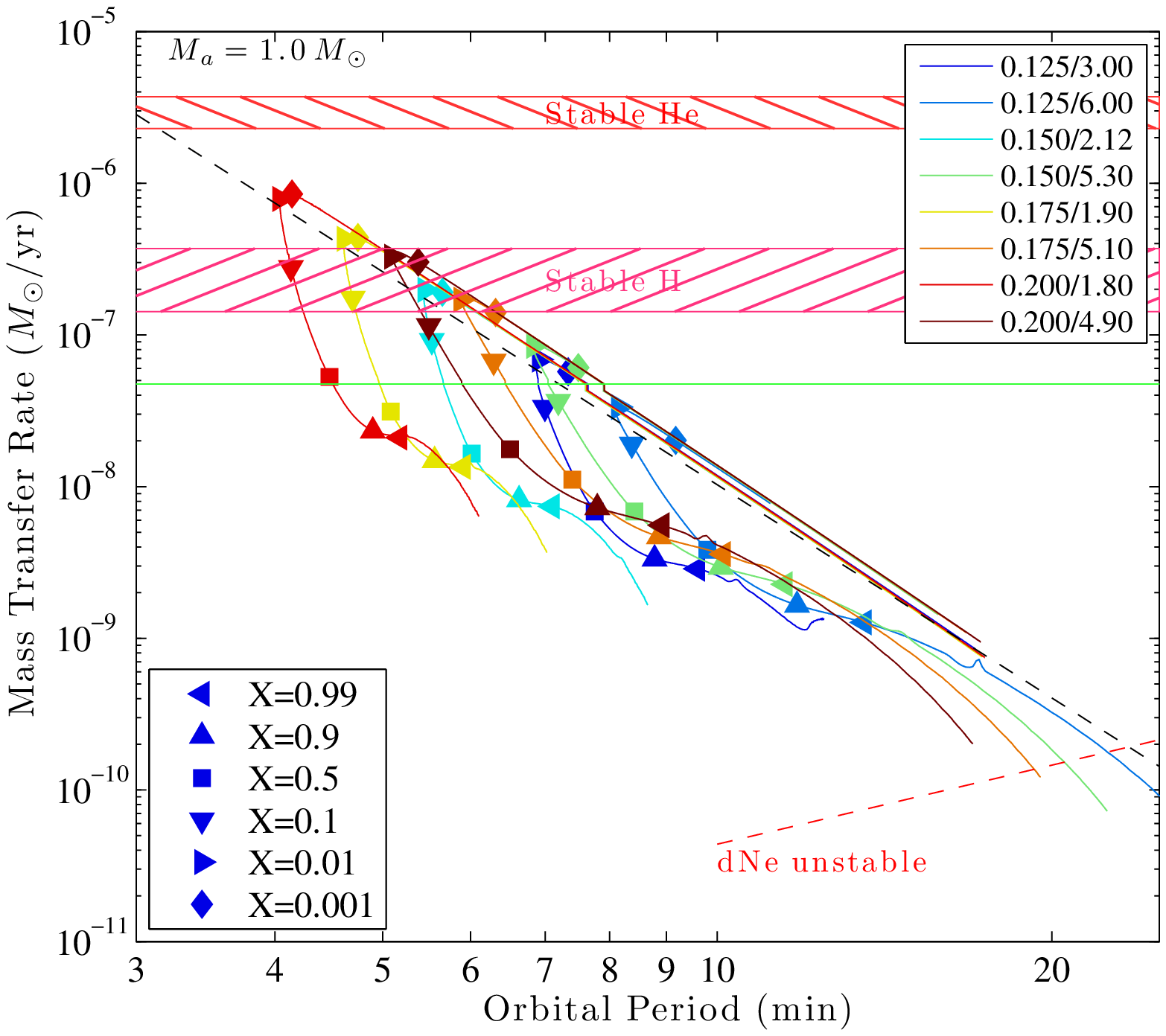}}
\caption{Evolution of compact WD binaries involving an ELM He WD (as
 in Figure~\ref{fig:mdot}), for accretor masses of $0.7\,M_\odot$
 (upper left), $0.8\,M_\odot$
 (upper right), $0.9\,M_\odot$
 (lower left), $1.0\,M_\odot$
 (lower right).}
\label{fig:mdot2}
\end{figure*}

After the initial evolution the degenerate portions of the donor come
to the surface and a minimum period is reached.  This minimum is near
where $|\dot M|$ is maximum, where direct impact often starts (forcing
us to adjust $\dot J$ accordingly), and where the mass transfer can in
principle become dynamically unstable.  For less massive accretors
direct-impact accretion will always happen.  The mass-transfer rate is
shown in Figures~\ref{fig:mdot} and \ref{fig:mdot2} for a single
accretor mass and a range of donor masses, while Figure~\ref{fig:pdot}
shows the orbital period derivative.  The mass-transfer factor $f$ is
calculated according to \S~\ref{sec:f}.

The interacting binary star HM~Cancri (\hmcnc\ or \rxj) has an orbital
period of 5.4\,min (\citealt{ipc+99}; \citealt*{rhc02};
\citealt{ihc+02}). Spectroscopy \citep{rrm+10} supports a model with
two interacting white dwarfs consistent with masses of $0.27\,M_\odot$
and $0.55\,M_\odot$ but with the actual values unconstrained.  X-ray
observations show a decreasing orbital period (\citealt{strohmayer05})
at a rate consistent with gravitational-wave emission, as shown in
Figure ~\ref{fig:pdot} (where we highlight it's value of $P$ and $\dot
P$).  While the X-ray luminosity was puzzlingly low for accretion from
a degenerate companion, \citet{davb+06} showed that the implied
transfer rate was much more consistent with an ELM WD donor.  We find
the same conclusion, which is that \hmcnc\ is consistent with
accretion from an ELM WD that is still inspiraling and transferring
H-rich material.  Such a system, along with other short-period
binaries like V407~Vul (RX J1914.4+2456;
\citealt{cham+98}), are extremely valuable as probes of the
eventual fates of systems like those in Figure~\ref{fig:wdmerg}.

\subsection{Impact on AM CVn formation} 
Differentiating between the different possible outcomes based on the
progenitors characteristics is a major goal of work like that
described here.  In particular, we seek to separate the systems that
will remain separate from those that will merge.  Systems that remain
stable will move to longer orbital periods with declining accretion
rates and become AM CVn binaries: systems that have a prolonged period
of He mass transfer driven by gravitational wave emission.  In
contrast, those that become unstable will shred the He WD into a
rapidly rotating hot disk that likely becomes an R~CrB star
\citep{sj02}.  \citet{mns04} did this in detail by considering the
stability of mass transfer, including the unknown effects of tidal
coupling.

Based on the qualitative effects described
in \S~\ref{sec:stable}, we can examine here whether more systems will
remain stable.  This will directly influence population calculations
\citep[e.g.,][]{bkapk11}, which find only a small fraction of the
total AM~CVn population could come from ELM WD binaries, although
based on the limited parameter space that we explore we do not make
population predictions in this work.

\begin{figure}
%\plotone{marsh_fmassX.eps}
\plotone{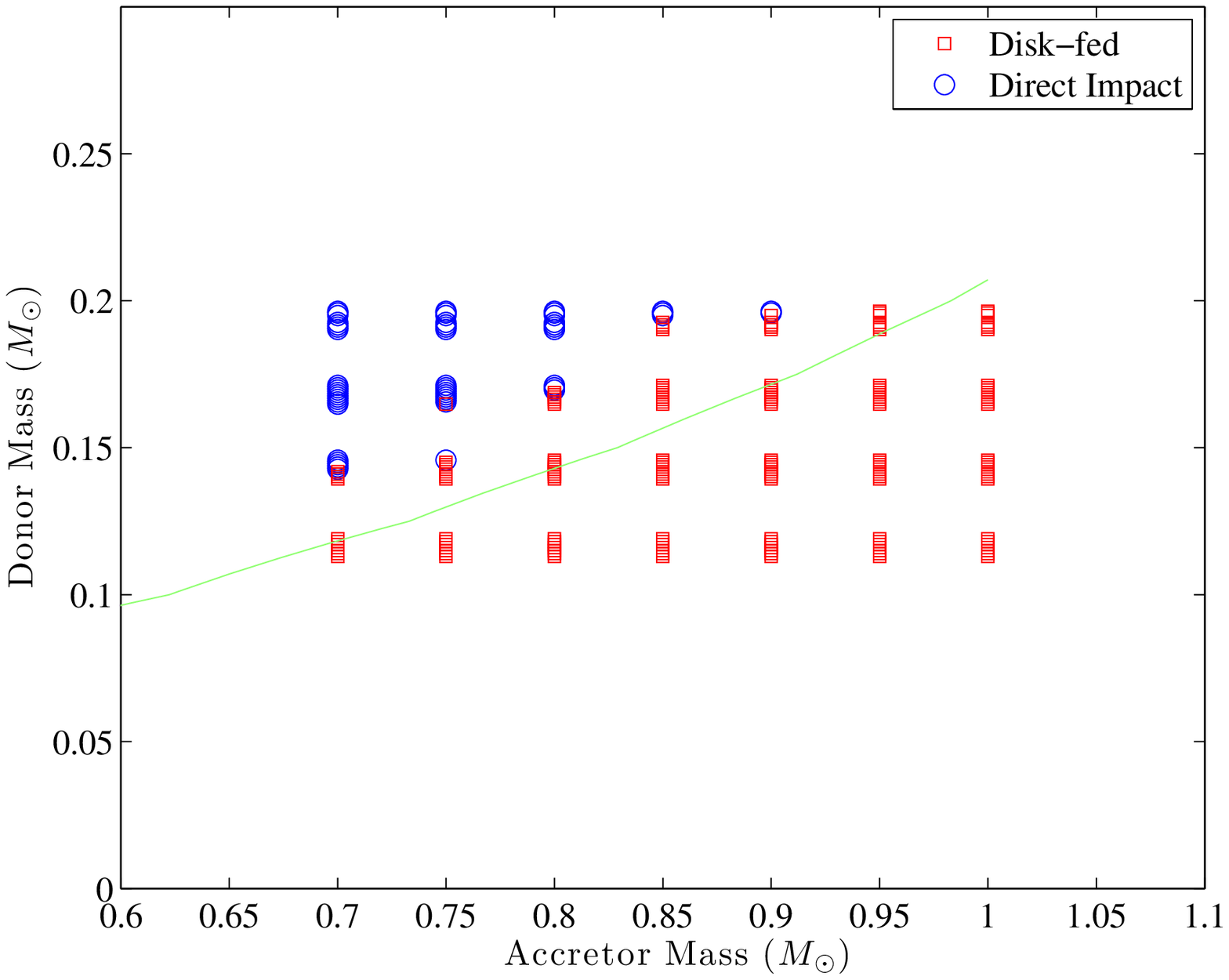}
\caption{Stability of mass transfer for different values of the donor
  and accretor mass, following \citet[][Fig.~1]{mns04}.  For each of
  our model calculations (some of which are shown in
  Fig.~\ref{fig:mdot2}), we plot a square if the accretion is always
  disk-fed (and hence guaranteed stable) and a circle if it enters a
  direct-impact period.  For each combination of donor and accretor
  mass, the different models correspond to different envelope masses
  that have been slightly offset vertically, for clarity.  The
  diagonal line is the direct/disk line from \citet{mns04}, using
  their cold equation-of-state for the donor.}
\label{fig:mns}
\end{figure}

In Figure \ref{fig:mns} we differentiate between those systems that
has disk-fed accretion and those that undergo direct impact.  For all
of the systems that we study dynamical stability
(Eqns.~\ref{eqn:stable} or \ref{eqn:stabledirect}) is maintained.  We
also see that  the increased size and decreased degeneracy of the
donor which begins mass transfer at larger separations means that more
space is available for accretion disks.  The exact threshold depends
on both the donor's core and envelope  masses, although in general it
increases by about $0.03\,M_\odot$ or so compared to the threshold for
a cold donor in \citet{mns04}.  

While none of the systems we study is dynamically unstable, this is
largely because we do not include the less massive accretors
(0.4--0.6\,$M_\odot$) which appear common \citep{kbap+12} and which
would be closer to the stability threshold.  Extending our calculation
to lower accretor masses with the addition of improved stable-burning
thresholds will help investigate dynamical stability more closely, but
the qualitative behavior of an increased range of systems maintaining
accretion disks and remaining stable will hold.  Current estimates,
based on rather small samples of ELM WDs and AM~CVns, suggest that
only a small fraction of AM~CVns come from ELM WDs \citep{bkapk11},
but significant improvements in the AM~CVn population
\citep[e.g.,][]{lfg+11}  and evolutionary calculations may change
this conclusion.  It is clear that this increased phase space for
stable accretion outcomes demonstrated here will lead to a higher
yield of AM CVn binaries from double WDs.

\subsection{Stable Outgoing Helium Transfer} 
\label{sec:outgoing}
The final behavior is familiar and the change in $J$ is
determined not by the change in $a$ but by the mass transferred.  The
mass-radius relation of the donor determines the evolution. In
this range, $\zhe$ is roughly constant (the classical value is $-1/3$,
and in our model is it close to $-1/4$).  With that, we find
\begin{equation}
P\approx \frac{9\pi}{\sqrt{2G}} \left[ R_0
 \left(\frac{\Mhe}{M_0}\right)^{\zeta_0} \Mhe^{-1/3}\right]^{3/2}
\end{equation}
where $\Rhe=R_0(\Mhe/M_0)^{\zeta_0}$, and $\zeta_0$ is the final
constant value of $\zhe$.  With this, we ignore the terms in the
brackets in Eqn.~(\ref{eqn:jdot}) that contain $\Mhe$ since they are
less than the other terms ($5/6+\zhe/2$), and find:
\begin{equation}
|\dot \Mhe|_{\rm out}\approx \frac{7776}{5} \frac{M_0 M_a
 R_0^3}{c^5}\left(\frac{4\pi^{14}G^2}{M_t P^{14}}\right)^{1/3}
\label{eqn:mdotporb}
\end{equation}
in the limit that $\zeta_0\approx -1/3$.  With $\zeta_0=-1/4$, the
exponent on $P$ changes from $-14/3=4.67$ to $-104/21=4.95$.

Based on these, we can determine
$\dot P$ along the incoming and outgoing branches:
\begin{eqnarray}
\dot P_{\rm in}&=& \frac{-384\Mhe}{5 c^5} \left(\frac{4\pi^8G^5
 M_a^2}{P^5}\right)^{1/3} \nonumber \\
\dot P_{\rm out}&=&\frac{1728}{5 c^5} \left(\frac{2 \pi^{22} G^7 R_0^9 M_0^3
 M_a^4}{P^{16}}\right)^{1/6}
\label{eqn:pdot}
\end{eqnarray}
This then allows us to determine the relative populations of objects
along the incoming and outgoing branches.  This is just the ratio of
the $\dot P$'s, since the time spent ($\propto 1/\dot P$) is related to the population:
\begin{equation}
{\cal N}_{\rm out/in}\equiv\left|\frac{\dot P_{\rm in}}{\dot P_{\rm out}}\right|\approx \sqrt{\frac{8G}{M_0}}\frac{P \Mhe}{9\pi R_0^{3/2}}.
\end{equation}
Based on our model, $R_0=0.034R_\odot$ at $M_0=0.05M_{\odot}$, so
${\cal N}_{\rm out/in}\approx 4 (P/{\rm
  10\,min})({\Mhe}/0.15\,M_\odot)$ where the mass of the He WD is the
original mass along the incoming track.  This shows that if every AM
CVn had an ELM WD as the originating double WD binary, then there is
one inspiraling H mass transferring system for every 4 AM CVns at an
orbital period of 10 minutes.  Note that this is not true in the
period/mass range where direct-impact accretion occurs (as presumably
for HM~Cnc), as that dramatically increases $|\dot P_{\rm out}|$ and
so reduces the outgoing (AM~CVn) population.  In any case, finding
these predicted systems remains a challenge.

\citet{davb+06} found it unlikely that V407~Vul had the same origin as
\hmcnc, as the lifetimes on the incoming versus outgoing branches
would suggest that its orbital period should be increasing rather than
decreasing (\citealt{strohmayer02} find $\dot
P=-\expnt{2.6}{-12}\,{\rm s\,s}^{-1}$ at an orbital period of
9.5\,min).  We do not find quite such a strong preference, with ${\cal
  N}_{\rm out/in}\approx 2$--4 depending on the accretor mass (compared
to a ratio of about 5 in \citealt{davb+06}).  Still, the small orbital
period derivative of V407~Vul is somewhat difficult to explain, being
a factor of 2 smaller than the lowest value we find in
Figure~\ref{fig:pdot}.  Considering lower accretor masses may help
resolve this.

\section{Conclusions}
We have shown that the unique properties of ELM WDs --- large,
nondegenerate H-rich shells supported by stable H burning --- lead to
some new phenomena when mass transfer initiates in double-WD binaries.
There is a prolonged period of H-rich mass transfer at a low rate
during inspiral, with \hmcnc\ potentially being the prototype. The
change in the mass-radius relation for the donor creates an
intrinsically more stable binary that opens up additional phase space
for making stable He accreting binaries. This may increase the AM CVn
birthrate, potentially alleviating the apparent paucity of progenitor
systems.  Those AM CVns which emerged from this progenitor scenario
will also have a larger He core radius than expected from an initially
cold WD, thereby exhibiting a higher accretion rate at a fixed orbital
period than from a cold WD \citep{db03}.

Prior to the onset of mass transfer, the ELM WDs had Gyrs to undergo
diffusive settling and substantial burning of hydrogen.  That clearly
allows for the complete sedimentation of the heavier elements from the
outermost layers of the WD. Hence, the mass transferred will vary from
nearly pure H, to nearly pure He. As we discussed, much work remains
to more carefully calculate the thermonuclear outcomes from this mass
transfer. If more thermally stable, then these systems may become more
observationally detectable due to the higher luminosities. It is also
interesting to note the pronounced absence of heavy elements in the
x-ray spectra of \hmcnc\ \citep{strohmayer08}, also pointing to an
ELM origin that lived a long time prior to mass transfer initiation.
If thermally unstable, then the accumulated mass could ignite
explosively \citep{bswn07}, potentially contributing to the increasing
number of low-luminosity ``supernovae'' observed locally \citep[e.g.,][]{kkgy+10}.

\acknowledgements We thank Bill Paxton for advice on running MESA, and
thank Bill Paxton and Ken Shen for helpful discussions.  This research
has been supported by the National Science Foundation under grants PHY
11-25915 and AST 11-09174.

%\bibliography{wd}

%% --------------------------------------------------------------------
%% Thu Jun 14 09:55:43 2012
%%   This file was generated automagically from the files
%%   evolution.bbl and evolution.tex using
%%     /Users/dlk//perl/nat2jour.pl
%%   This file should accompany evolution-aas.tex.
%% --------------------------------------------------------------------

\end{document}